\documentclass
[preprint,byrevtex,nofootinbib,10pt,balancelastpage,titlepage,bibnotes,prb,twocolumn]{revtex4}%
\usepackage{amsfonts}
\usepackage{amsmath}
\usepackage{amssymb}
\usepackage{graphicx}%
\providecommand{\U}[1]{\protect\rule{.1in}{.1in}}
\ifx\pdfoutput\relax\let\pdfoutput=\undefined\fi
\newcount\msipdfoutput
\ifx\pdfoutput\undefined\else
\ifcase\pdfoutput\else
\ifx\paperwidth\undefined\else
\ifdim\paperheight=0pt\relax\else\pdfpageheight\paperheight\fi
\ifdim\paperwidth=0pt\relax\else\pdfpagewidth\paperwidth\fi
\fi\fi\fi
\begin{document}
\preprint{UATP/1301}
\title{Nonequilibrium Entropy}
\author{P.D. Gujrati}
\email{pdg@uakron.edu}
\affiliation{Department of Physics, Department of Polymer Science, The University of Akron,
Akron, OH 44325}

\begin{abstract}
We consider an isolated system in an arbitrary state and provide a general
formulation using first principles for an \emph{additive} and
\emph{non-negative} statistical quantity $\mathcal{S}_{0}(t)$ that is shown to
reproduce the equilibrium thermodynamic entropy of the isolated system. We
further show that $\mathcal{S}_{0}(t)$ represents the nonequilibrium
thermodynamic entropy $S_{0}(t)$ when the latter is a state function of
nonequilibrium state variables; see text. We consider an isolated $1$-d ideal
gas and determine its non-equilibrium statistical entropy $\mathcal{S}_{0}(t)$
as a function of the box size as the gas expands freely isoenergetically, and
compare it with the equilibrium thermodynamic entropy $S_{0\text{eq}}$. We
find that $\mathcal{S}_{0}(t)\leq$ $S_{0\text{eq}}$ in accordance with the
second law, as expected. To understand how $\mathcal{S}_{0}(t)$ is different
from thermodynamic entropy of classical continuum models that is known to
become\emph{ negative} under certain conditions, we calculate $\mathcal{S}%
_{0}\ $for a $1$-d lattice model and discover that it can be related to the
thermodynamic entropy of the continuum $1$-d Tonks gas by taking the lattice
spacing $\delta$ go to zero. However, $\Delta\mathcal{S}_{0}\neq$ $\Delta
S_{0}$ since $\delta$ is state-dependent. We discuss the semi-classical
approximation of our entropy and show that the standard quantity $S_{f}(t)$ in
the Boltzmann's H-theorem, see Eq. (\ref{Sf_S}), does not directly correspond
to the statistical entropy.

\end{abstract}
\date{\today}
\maketitle

\section{Introduction}

\begin{figure}
[ptb]
\begin{center}
\includegraphics[
height=1.772in,
width=3.4584in
]%
{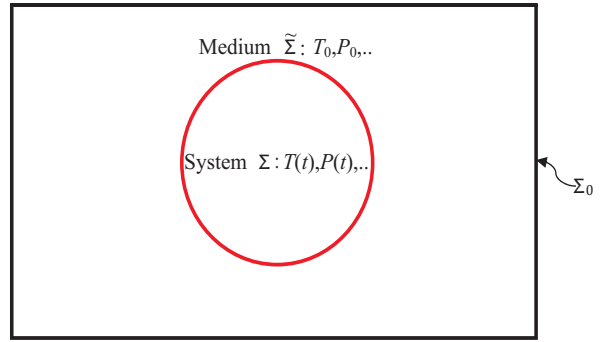}%
\caption{Schematic representation of $\Sigma$, $\widetilde{\Sigma}$ and
$\Sigma_{0}$. We assume that $\Sigma$ and $\widetilde{\Sigma}$ are homogeneous
and in internal equilibrium, but not in equilibrium with each other. The
internal fields $T(t),P(t),\cdots$ fof $\Sigma$ and $T_{0},P_{0},\cdots$ of
$\widetilde{\Sigma}$ are not the same unless they are in equilibrium with each
other. There will be viscous dissipation in $\Sigma$ when not in equilibrium
with $\widetilde{\Sigma}$.}%
\label{Fig.Sys}%
\end{center}
\end{figure}

Although the concept of entropy plays important roles in diverse fields
ranging from classical thermodynamics of Clausius,
\cite{Clausius,Gibbs,Rice,Tolman,Landau} quantum mechanics and uncertainty,
\cite{von Neumann,Landau-QM,Partovi} black holes, \cite{Beckenstein} coding
and computation,\cite{Schumacher,Bennet} to information technology,
\cite{Wiener,Shannon} it does not seem to have a standard definition. Here, we
are interested in its application to nonequilibrium statistical
thermodynamics. In classical thermodynamics, it is defined as a
\emph{thermodynamic }quantity with no association of any notion of a
microstate and its probability, while modern approach to\ statistical
thermodynamics, primarily due to Boltzmann and Gibbs, requires a probabilistic
approach in terms of microstates. In this work, we are primarily interested in
an isolated system $\Sigma_{0}$. Quantities for the isolated system will carry
a suffix $0$; quantities without the suffix will refer to any body, which need
not be isolated, such as a part $\Sigma$ of $\Sigma_{0}$; see Fig.
\ref{Fig.Sys}. The microstates for $\Sigma_{0}$ are determined by the set
$\mathbf{X}_{0}$ of extensive observables (the energy $E_{0}$, the volume
$V_{0}$, the number of particles $N_{0}$ , etc.) specifying the isolated
system. While temporal evolution is not our primary interest in this work, we
still need to remember the importance of temporal evolution in thermodynamics.
We will say that two microstates belonging to the microstate subspace
$\Gamma_{0}(\mathbf{X}_{0})$ are "connected" if one evolves from the other
after some time $\tau_{0}<\infty$. Before this time, they will be treated as
"disconnected." Let $\tau_{0\text{max}}$ denote the maximum $\tau_{0}$ over
all pairs of microstates. The space $\Gamma_{0}(\mathbf{X}_{0})$ is
\emph{simply connected} for all times longer than $\tau_{0\text{max}}$ in that
each microstate can evolve into another microstate $\in\Gamma_{0}%
(\mathbf{X}_{0})$ in due time. For $t<\tau_{0\text{max}}$, the space
$\Gamma_{0}(\mathbf{X}_{0})$ will consist of \emph{disjoint components,} an
issue that neither Boltzmann nor Gibbs has considered to the best of our
knowledge. But the issue, which we consider later in Sec.
\ref{Marker_Disjoint Space}, becomes important in considering nonequilibrium
states.%
%TCIMACRO{\FRAME{ftbpFU}{3.4584in}{1.772in}{0pt}{\Qcb{Schematic representation
%of $\Sigma$, $\widetilde{\Sigma}$ and $\Sigma_{0}$. We assume that $\Sigma$
%and $\widetilde{\Sigma}$ are homogeneous and in internal equilibrium, but not
%in equilibrium with each other. The internal fields $T(t),P(t),\cdots$ fof
%$\Sigma$ and $T_{0},P_{0},\cdots$ of $\widetilde{\Sigma}$ are not the same
%unless they are in equilibrium with each other. There will be viscous
%dissipation in $\Sigma$ when not in equilibrium with $\widetilde{\Sigma}$.}%
%}{\Qlb{Fig.Sys}}{system_modified_1.eps}{\special{ language "Scientific Word";
%type "GRAPHIC";  maintain-aspect-ratio TRUE;  display "USEDEF";
%valid_file "F";  width 3.4584in;  height 1.772in;  depth 0pt;
%original-width 5.1673in;  original-height 2.4915in;  cropleft "0.0775";
%croptop "1";  cropright "1.0258";  cropbottom "0";
%filename 'System_Modified_1.eps';file-properties "XNPEU";}}}%
%BeginExpansion
%EndExpansion

Boltzmann assumes \emph{equal probability} of various microstates in the
simply connected set $\Gamma_{0}(\mathbf{X}_{0})$.\ Thus, $\tau_{0\text{max}}$
can be identified with the equilibration time $\tau_{0\text{eq}}%
(\mathbf{X}_{0})$ for $\Sigma_{0}$.\ Under the equiprobable assumption,
Boltzmann identifies the entropy in terms of the number of
microstates\cite{Planck,Landau} $W_{0}\left(  \mathbf{X}_{0}\right)  $ in
$\Gamma_{0}\left(  \mathbf{X}_{0}\right)  $:
\begin{equation}
S_{0\text{B}}\left(  \mathbf{X}_{0}\right)  \equiv\ln W_{0}\left(
\mathbf{X}_{0}\right)  \text{;} \label{Boltzmann_S}%
\end{equation}
we will set the Boltzmann constant to be unity throughout the work so that the
entropy will always be a \emph{pure number}. The idea behind the above formula
implicitly appears for the first time in a paper \cite{Boltzmann} by
Boltzmann, and then appears more or less in the above form later in his
lectures\cite{Boltzmann0} where he introduces the combinatorial approach for
the first time to statistical mechanics. The formula itself does not appear
but is implied when he takes the logarithm of the number of
combinations.\cite{Boltzmann0,Note-Boltzmann} (There is another formulation
for entropy given by Boltzmann,\cite{Boltzmann0,Boltzmann} which is also known
as the Boltzmann entropy,\cite{Jaynes} that we will discuss later and that has
a restricted validity; see Eq. (\ref{Boltzmann_S_1}).) Gibbs, also using the
probabilistic approach, gives the following formula for the entropy in a
\emph{canonical ensemble}:\cite{Gibbs,Landau}%
\begin{equation}
S_{\text{G}}^{(\text{c})}\equiv-%
%TCIMACRO{\tsum \limits_{\alpha\in\Gamma_{\text{E}}}}%
%BeginExpansion
{\textstyle\sum\limits_{\alpha\in\Gamma_{\text{E}}}}
%EndExpansion
p_{\alpha}^{(\text{c})}\ln p_{\alpha}^{(\text{c})};\ \
%TCIMACRO{\tsum \limits_{\alpha\in\Gamma_{\text{E}}}}%
%BeginExpansion
{\textstyle\sum\limits_{\alpha\in\Gamma_{\text{E}}}}
%EndExpansion
p_{\alpha}^{(\text{c})}=1 \label{Gibbs_S}%
\end{equation}
where $p_{\alpha}^{(\text{c})}$ is the canonical ensemble probability of the
$\alpha$th microstate of the system, and the sum is over all microstates
corresponding to all possible energies $E$ (with other elements in
$\mathbf{X}$ held fixed); their set is denoted by $\Gamma_{\text{E}}$ in the
above sum. The Gibbsian approach assumes an ensemble at a given instant, while
the Boltzmann approach considers the evolution of a particular system in time;
see for example a recent review.\cite{Gujrati-Symmetry} In equilibrium, both
entropy expressions yield the same result. In quantum mechanics, this entropy
is given by the von Neumann entropy formulation\cite{von Neumann,Landau-QM} in
terms of the density matrix $\rho$:%
\[
S_{\text{vN}}=-Tr(\rho\ln\rho).
\]
The entropy formulation in the information theory\cite{Wiener,Shannon} has a
form that appears to be similar in form to the above Gibbs entropy even though
the temperature has no significance in the information theory. There is also
another statistical formulation of entropy, heavily used in the literature, in
terms of the phase space distribution function $f(x,t)$, which follows from
Boltzmann's celebrated H-theorem:%
\begin{equation}
S_{f}(t)=-%
%TCIMACRO{\tint }%
%BeginExpansion
{\textstyle\int}
%EndExpansion
f(x,t)\ln f(x,t)dx; \label{Sf_S}%
\end{equation}
here, $x$ denotes a point in the phase space. This quantity is not only not
dimensionless but, as we will show later, is not the correct formulation in
general; see Eq. (\ref{Discrete_S}).

The classical thermodynamics entropy $S_{0}$ is oblivious to the microstates
and their probabilities and deals with $E_{0},V_{0},N_{0}$, etc. as the
observables of the system. In equilibrium, the entropy $S_{0}~$is a state
function, and can be expressed as a function of the observables. This
functional dependence results in the Gibbs fundamental relation
\begin{equation}
dS_{0}=%
%TCIMACRO{\tsum \nolimits_{p}}%
%BeginExpansion
{\textstyle\sum\nolimits_{p}}
%EndExpansion
\left(  \partial S_{0}/\partial X_{0p}\right)  dX_{0p}
\label{Gibbs_Fundamental}%
\end{equation}
in terms of the observables $\left\{  X_{0p}\right\}  $. For a lattice model,
$S_{0}$ is non-negative in accordance with the Boltzmann definition of
$S_{0\text{B}}\left(  \mathbf{X}_{0}\right)  $, but is known to become
negative for a continuum model such as for an ideal gas. The latter
observation implies that such continuum models are not realistic as they
violate Nernst's postulate(the third law) and require \emph{quantum mechanics}
to ensure non-negativity of the entropy.\cite{Landau} Even the change $\Delta
S_{0}$, the heat capacity, etc. do not satisfy thermodynamic consequences of
Nernst's postulate.

By invoking Nernst's postulate (the equilibrium entropy vanishes at absolute
zero), one can determine the equilibrium entropy everywhere \emph{uniquely}.
The consensus is that in equilibrium, the thermodynamic entropy is not
different from the above statistical entropies due to Boltzmann and Gibbs.
However, there is at present no consensus when the system is out of
equilibrium. There is also some doubt whether the nonequilibrium thermodynamic
entropy has any meaning in classical thermodynamics. We will follow
Clausius\ and take the view here that the thermodynamic entropy is a
well-defined notion even for an irreversible process going on in a body for
which Clausius\cite[p. 214]{Clausius} writes $TdS>d_{\text{e}}Q$ in terms of
the exchange heat $d_{\text{e}}Q$ with the medium. The question that arises is
whether the two statistical definitions can be applied to a body out of
equilibrium. We find the answer to be affirmative. The next question that
arises is the following: Do they always give the same results?\ We will show
that under certain conditions, they give the same results. This is important
as the literature is not very clear on this
issue.\cite{Lavis,Bishop,Lebowitz,Ruelle}\ 

For an isolated system, which we mostly consider here, we are not concerned
with any thermostat or external working source. As a consequence, observables
$E_{0}$,$V_{0}$,$N_{0}$, etc. in $\mathbf{X}_{0}$ must remain constant even if
the system is out of equilibrium. While this will simplify our discussion to
some extent, it will also create a problem as we will discuss later. We
discuss the concept of a nonequilibrium state, state variables and state
functions in the next section. We introduce the statistical entropy
formulation in Sec. III and show its equivalence with thermodynamic
nonequilibrium entropy when the latter is a state function. In Sec. IV, we
carry out an explicit quantum calculation of the nonequilibrium statistical
entropy.\ In Sec. V, we consider a $1$-d lattice model\ appropriate for Tonks
gas in continuum so that the statistical lattice entropy can be calculated
rigorously. We take the continuum limit and compare the resulting entropy with
the continuum entropy of the Tonks gas and obtain an interesting result. We
discuss the semi-classical approximation of the statistical entropy in Sec. VI
and show that the formulation in Eq. (\ref{Sf_S}) does not determine the
entropy. A brief summary and discussion is presented in the final section.

\section{The Second Law and A Nonequilibrium State}

\subsection{Second Law}

The second law states that the irreversible (denoted by a suffix i) entropy
generated in any infinitesimal physical process going on in a body satisfies
the inequality%
\begin{equation}
d_{\text{i}}S\geq0; \label{Second_Law0}%
\end{equation}
the equality occurs for a reversible process. For an isolated system, there is
no exchange (denoted by a suffix e) entropy change $d_{\text{e}}S_{0}$ with
the medium so that $dS_{0}=d_{\text{e}}S_{0}+d_{\text{i}}S_{0}$ in any
arbitrary process. in an isolated system satisfies%
\begin{equation}
dS_{0}=d_{\text{i}}S_{0}\geq0. \label{Second_Law}%
\end{equation}
The law refers to the \emph{thermodynamic entropy}. It is not a state function
in the conventional sense ($S_{0}(\mathbf{X}_{0})$) if $\Sigma_{0}$ is not in
equilibrium simply because as a state function $S_{0}(\mathbf{X}_{0})$ must
remain constant for constant $\mathbf{X}_{0}$; see below also.

As the thermodynamic entropy is not measurable except when the process is
reversible, the second law remains useless as a computational tool. In
particular, it says nothing about the rate at which the irreversible entropy
increases. Therefore, it is useful to obtain a computational formulation of
the entropy, the \emph{statistical entropy}. This will be done in the next
section. The onus is on us to demonstrate that the statistical entropy also
satisfies this law if it is to represent the thermodynamic entropy. This by
itself does not prove that the two are the same. It is not been possible to
show that the statistical entropy is identical to the thermodynamic entropy in
general. Here, we show their equivalence only when the nonequilibrium
thermodynamic entropy is a \emph{state function} of nonequilibrium state
variables to be introduced below.

\subsection{Concept of a Nonequilibrium
State\label{Marker_Internal Equilibrium}}

For an isolated body in equilibrium, the entropy can be expressed as a
function of its observables (variables that can be controlled by an observer),
as is easily seen form the Gibbs fundamental relation in Eq.
(\ref{Gibbs_Fundamental}). The thermodynamic state, also known as the
macrostate $\mathcal{M}_{0},$ of a body in equilibrium remains the same unless
it is disturbed. Therefore, we can identify the equilibrium state
$\mathcal{M}_{0\text{eq}}$ of the body by its observables. Accordingly, its
equilibrium entropy $S_{0}(\mathbf{X}_{0})$ can be expressed as a function of
its observables $\mathbf{X}_{0}$. This is expressed by saying that the
equilibrium entropy is a \emph{state function} and $\mathbf{X}_{0}$\ is the
set of \emph{state variables}.

The above conclusion is most certainly not valid for a body out of
equilibrium, which\ we take to be isolated. If the body is not in equilibrium,
its (macro)state $\mathcal{M}_{0}(t)$ will continuously change, which is
reflected in its entropy change (increase) in time; this requires expressing
its entropy as $S_{0}(\mathbf{X}_{0},t)$ with an \emph{explicit
time-dependence}, since $\mathbf{X}_{0}=constant$ for an isolated body. The
change in the entropy and the macrostate must come from the variations of
additional variables, distinct from the observables, that keep changing with
time until the body comes to equilibrium as explained
elsewhere.\cite{Gujrati-I,Gujrati-II} These variables cannot be controlled by
the observer. Once the body has come to equilibrium, the entropy has no
explicit time-dependence and becomes a state function. In this state, the
entropy has its maximum possible value for given $\mathbf{X}_{0}$. In other
words, when the entropy becomes a state function, it achieves the maximum
possible value for the given set of state variables, here given by
$\mathbf{X}_{0}$. This conclusion about the entropy will play an important
role below.

We assume that there is a set $\boldsymbol{\xi}_{0}$ of\ additional variables,
known as the \emph{internal variables }(sometimes also called hidden
variables). We will refer to the variables in $\mathbf{X}_{0}$ and
$\boldsymbol{\xi}_{0}$\ as (\emph{nonequilibrium})\ \emph{state variables}
(see below for justification) and denote them collectively as $\mathbf{Z}_{0}$
in the following. From Theorem 4 presented elsewhere,\cite{Gujrati-II} it
follows that with a proper choice of the number of internal variables, the
entropy can be written as $S_{0}(\mathbf{Z}_{0}(t))$ with no explicit
$t$-dependence. The situation is now almost identical to that of an isolated
body in equilibrium:\ The entropy is a function of $\mathbf{Z}_{0}(t)$ with no
explicit time-dependence. This allows us to identify $\mathbf{Z}_{0}(t)$\ as
the set of nonequilibrium state variables. Thus, $\mathcal{M}_{0}(t)$ can be
specified by $\mathbf{Z}_{0}(t)$ so that the entropy becomes a \emph{state
function}. This allows us to extend Eq. (\ref{Gibbs_Fundamental}) to%
\begin{equation}
dS_{0}=%
%TCIMACRO{\tsum \nolimits_{p}}%
%BeginExpansion
{\textstyle\sum\nolimits_{p}}
%EndExpansion
\left(  \partial S_{0}/\partial Z_{0p}\right)  dZ_{0p}
\label{Gibbs_Fundamental_Extended}%
\end{equation}
in which the partial derivatives are related to the fields of the system:%
\begin{equation}
\left(  \partial S_{0}/\partial E_{0}\right)  =1/T_{0},\left(  \partial
S_{0}/\partial V_{0}\right)  =P_{0}/T_{0},\cdots\label{Fields_Isolated}%
\end{equation}
these fields will change in time unless the system has reached equilibrium.

As $\mathbf{Z}_{0}(t)$\ changes in time, $\mathcal{M}_{0}(t)$ changes, but at
each instant the (nonequilibrium) entropy as a state function, has a maximum
possible value for given $\mathbf{Z}_{0}(t)$ even though $\mathcal{M}%
_{0}(t)\neq\mathcal{M}_{0\text{eq}}$. In our previous
work,\cite{Gujrati-I,Gujrati-II,Gujrati-III} we have identified this
particular state as an \emph{internal equilibrium state}, but its physical
significance as presented above was not discussed. For a state that is not in
internal equilibrium, the entropy must retain an explicit time-dependence. In
this case, the derivatives in Eq. (\ref{Fields_Isolated}) cannot be identified
as state variables like, temperature, pressure, etc.

It may appear to a reader that the concept of entropy being a state function
is very restrictive. This is not the case as this concept, although not
recognized by several workers, is implicit in the literature where the
relationship of the thermodynamic entropy with state variables is
investigated. To appreciate this, we observe that the entropy of a body in
internal equilibrium\cite{Gujrati-I,Gujrati-II} is given by the Boltzmann
formula%
\begin{equation}
S(\mathbf{Z}(t))=\ln W(\mathbf{Z}(t)), \label{Boltzmann_S_Extended}%
\end{equation}
in terms of the number of microstates corresponding to $\mathbf{Z}(t)$. In
classical nonequilibrium thermodynamics,\cite{deGroot} the entropy is always
taken to be a state function. In the Edwards approach\cite{Edwards} for
granular materials, all microstates are equally probable as is required for
the above Boltzmann formula. Bouchbinder and Langer\cite{Langer} assume that
the nonequilibrium entropy is given by Eq. (\ref{Boltzmann_S_Extended}).
Lebowitz\cite{Lebowitz} also takes the above formulation for his definition of
the nonequilibrium entropy. As a matter of fact, we are not aware of any work
dealing with entropy computation that does not assume the nonequilibrium
entropy to be \ a state function. This does not, of course, mean that all
states of a system are internal equilibrium states. For states that are not in
internal equilibrium, the entropy is not a state function so that it will have
an explicit time dependence. But, as shown elsewhere,\cite{Gujrati-II} this
can be avoided by enlarging the space of internal variables. The choice of how
many internal variables are needed will depend on experimental time scales and
cannot be answered in generality at present. We hope to come back to this
issue in a future publication.

For a general body that is not isolated, the concept of its internal
equilibrium state plays a very important role in that the body can come back
to this state several times in a nonequilibrium process. In a cyclic
nonequilibrium process, such a state can repeat itself in time after some
cycle time $\tau_{\text{c}}$ so that all state variables and functions
including the entropy repeat themselves:%
\[
\mathbf{Z}(t+\tau_{\text{c}})=\mathbf{Z}(t),~\mathcal{M}(t+\tau_{\text{c}%
})=\mathcal{M}(t),~S(t+\tau_{\text{c}})=S(t).
\]
This ensures $\Delta_{\text{c}}S\equiv S(t+\tau_{\text{c}})-S(t)=0$ in a
cyclic process. All that is required for the cyclic process to occur is that
the body must start and end in the same internal equilibrium state; however,
during the remainder of the cycle, the body need not be in internal equilibrium.\ 

\section{General Formulation of the Statistical Entropy\label{Marker_NonEq-S}%
\ \ \ }

We provide a very general formulation of the statistical entropy, which will
also demonstrate that the entropy is a \emph{statistical average}. We consider
a macrostate $\mathcal{M}_{0}(t)\equiv\mathcal{M}_{0}(\mathbf{Z}_{0}(t))$ of
$\Sigma_{0}$\ at a given instant $t$. In the following, we suppress $t$ unless
necessary. The macrostate $\mathcal{M}_{0}(\mathbf{Z}_{0})$ refers to the set
of microstates $\mathbf{m}_{0}=\left\{  m_{0\alpha}\right\}  $\ and their
probabilities $\mathbf{p}=\left\{  p_{\alpha}\right\}  $. For the computation
of combinatorics, the probabilities are handled in the following abstract way.
We consider a large number $\mathcal{N}_{0}\mathcal{=C}_{0}W_{0}%
(\mathbf{Z}_{0})$ of independent \emph{replicas} or \emph{samples} of
$\Sigma_{0}$, with $\mathcal{C}_{0}$\ some large constant integer and
$W_{0}(\mathbf{Z}_{0})$\ the number of distinct microstates $m_{0\alpha}$. The
samples should be thought of as identically prepared experimental
samples.\cite{Gujrati-Symmetry} Let $\Gamma_{0}(\mathbf{Z}_{0})$ denote the
sample space spanned by $m_{0\alpha}$.

\subsection{Simply Connected Sample
Space\ \ \ \ \ \ \ \ \ \ \ \ \ \ \ \ \ \ \ \ \ \ \ \ \ \ \ \ \ \ \ \ \ \ \ \ \ \ \ \ \ \ \ \ \ \ \ \ \ \ \ \ \ \ \ \ \ \ \ \ \ \ \ \ \ \ \ \ \ \ \ \ \ \ \ \ \ \ \ \ \ \ \ \ \ \ \ \ \ \ \ \ \ \ \ \ \ \ \ \ \ \ \ \ \ \ \ \ \ \ \ \ \ \ \ \ \ \ \ \ \ \ \ \ \ \ \ \ \ \ \ \ \ \ \ \ \ \ \ \ \ \ \ \ \ \ \ \ \ \ \ \ \ \ \ \ \ \ \ \ }%

\subsubsection{An Isolated System\label{Marker_Probabilities}}

As $\Gamma_{0}(\mathbf{Z}_{0})\subset$ $\Gamma_{0}(\mathbf{X}_{0})$,
$\Gamma_{0}(\mathbf{Z}_{0})$\ is simply connected if $\Gamma_{0}%
(\mathbf{X}_{0})$ is, which we assume in this section.\ Let $\mathcal{N}%
_{0\alpha}(t)$ denote the number of $m_{0\alpha}$-samples (samples in the
$m_{0\alpha}$-microstate) so that%
\begin{equation}
0\leq p_{\alpha}(t)=\mathcal{N}_{0\alpha}(t)/\mathcal{N}_{0}\leq1;\ \
%TCIMACRO{\tsum \limits_{\alpha=1}^{W_{0}(\mathbf{Z}_{0})}}%
%BeginExpansion
{\textstyle\sum\limits_{\alpha=1}^{W_{0}(\mathbf{Z}_{0})}}
%EndExpansion
\mathcal{N}_{0\alpha}(t)=\mathcal{N}_{0}. \label{sample_probability}%
\end{equation}
The above sample space is a generalization of the \emph{ensemble} introduced
by Gibbs, except that the latter is restricted to an equilibrium system,
whereas our sample space refers to the system in any arbitrary state so that
$p_{\alpha}$ may be time-dependent. In the semi-classical approximation, see
Sec. \ref{Marker_SemiClassical}, one can similarly take the sample space to
represent the classical phase space of Boltzmann. The (\emph{sample }or\emph{
ensemble})\emph{ average} of some quantity $Q_{0}$\ over these samples is
given by%
\begin{equation}
\overline{Q_{0}}\equiv%
%TCIMACRO{\tsum \limits_{\alpha=1}^{W_{0}(\mathbf{Z}_{0})}}%
%BeginExpansion
{\textstyle\sum\limits_{\alpha=1}^{W_{0}(\mathbf{Z}_{0})}}
%EndExpansion
p_{\alpha}(t)Q_{0\alpha},\ \
%TCIMACRO{\tsum \limits_{\alpha=1}^{W_{0}(\mathbf{Z}_{0})}}%
%BeginExpansion
{\textstyle\sum\limits_{\alpha=1}^{W_{0}(\mathbf{Z}_{0})}}
%EndExpansion
p_{\alpha}(t)\equiv1, \label{Ensemble_Average}%
\end{equation}
where $Q_{0\alpha}$ is the value of $Q_{0}$ in $m_{0\alpha}$.

The samples are, by definition, \emph{independent} of each other so that there
are no correlations among them. Because of this, we can treat the samples to
be the outcomes of some random variable, the macrostate $\mathcal{M}_{0}(t)$.
This independence property of the outcomes is crucial in the following, and
does not imply that they are equiprobable. The number of ways $\mathcal{W}%
_{0}$ to arrange the $\mathcal{N}_{0}$ samples into $W_{0}(\mathbf{Z}_{0})$
distinct microstates is%
\begin{equation}
\mathcal{W}_{0}\mathcal{\equiv N}_{0}!/%
%TCIMACRO{\tprod \limits_{\alpha}}%
%BeginExpansion
{\textstyle\prod\limits_{\alpha}}
%EndExpansion
\mathcal{N}_{0\alpha}(t)!. \label{Combinations}%
\end{equation}
Taking its natural log to obtain an \emph{additive} quantity per sample%
\begin{equation}
\mathcal{S}_{0}\equiv\ln\mathcal{W}_{0}/\mathcal{N}_{0},
\label{Ensemble_entropy_Formulation}%
\end{equation}
and using Stirling's approximation, we see easily that $\mathcal{S}_{0}$,
which we hope to identify later with the entropy $S_{0}(\mathbf{Z}_{0}(t))$ of
the isolated system, can be written as the average of the negative of%
\[
\eta(t)\equiv\ln p(t),
\]
what Gibbs \cite{Gibbs} calls the \emph{index of probability:}
\begin{equation}
\mathcal{S}_{0}(\mathbf{Z}_{0}(t),t)\equiv-\overline{\eta(t)}\equiv-%
%TCIMACRO{\tsum \limits_{\alpha=1}^{W_{0}(\mathbf{Z}_{0})}}%
%BeginExpansion
{\textstyle\sum\limits_{\alpha=1}^{W_{0}(\mathbf{Z}_{0})}}
%EndExpansion
p_{\alpha}(t)\ln p_{\alpha}(t), \label{Gibbs_Formulation}%
\end{equation}
where we have also shown an explicit time-dependence for the reason that will
become clear below. The above derivation is based on fundamental principles
and does not require the system to be in equilibrium; therefore, it is always
applicable. To the best of our knowledge, even though such an expression has
been extensively used in the literature, it has been used \emph{without} any
derivation; one simply appeals to this form by invoking it as the information
entropy; however, see Sec. \ref{marker_Summary}.

Because of its similarity in form with $S_{\text{G}}^{(\text{c})}$ in Eq.
(\ref{Gibbs_S}), we will refer to $\mathcal{S}_{0}(\mathbf{Z}_{0}(t),t)$ as
the Gibbs statistical entropy from now on. As the nonequilibrium thermodynamic
entropy for a process in which the system is always in internal equilibrium
can be determined by integrating the Gibbs fundamental relation in Eq.
(\ref{Gibbs_Fundamental_Extended}), we can compare it with the statistical
entropy introduced above. However, such an integration is not possible for a
process involving states that are arbitrary (not in internal equilibrium).
Therefore, there is no meaning to compare $\mathcal{S}_{0}(\mathbf{Z}%
_{0}(t),t)$ with the corresponding thermodynamic entropy whose value cannot be
determined. To identify $\mathcal{S}_{0}(\mathbf{Z}_{0}(t))$ with the
nonequilibrium thermodynamic entropy requires the following additional steps:

\begin{enumerate}
\item[(1)] It is necessary to establish that $\mathcal{S}_{0}(t)$ satisfies
Eq. (\ref{Second_Law}).

\item[(2)] For an equilibrium canonical system, it is necessary to establish
that $\mathcal{S}(t)$ is identical to the equilibrium thermodynamic entropy
given by $S_{\text{G}}^{(\text{c})}$.\cite{Gibbs}

\item[(3)] It is necessary to show that $\mathcal{S}_{0}(t)$ is identical to
the nonequilibrium thermodynamic entropy of the system that is out of
equilibrium but whose entropy is a state function.
\end{enumerate}

There are several proofs available in the
literature\cite{Tolman,Rice,Jaynes,Gujrati-Residual,Gujrati-Symmetry} for (1).
Therefore, we will not be concerned with (1) anymore. We will prove (2) and
(3) in Sect. \ref{marker_Open system}.

The maximum possible value of $\mathcal{S}_{0}(t)$ for given $\mathbf{Z}%
_{0}(t)$ occurs when $m_{0\alpha}$ are \emph{equally probable}:%
\[
p_{\alpha}(t)\rightarrow p_{\alpha\text{,eq}}=1/W_{0}(\mathbf{Z}%
_{0}),\text{~~\ \ }\forall m_{\alpha}\in\Gamma_{0}(\mathbf{Z}_{0}).
\]
\ In this case, the explicit time dependence in $\mathcal{S}_{0}(t)$ will
\emph{disappear} and we have
\begin{equation}
\mathcal{S}_{0,\text{max}}(\mathbf{Z}_{0}(t),t)=\mathcal{S}_{0}(\mathbf{Z}%
_{0}(t))=\ln W_{0}(\mathbf{Z}_{0}(t)), \label{S_Boltzmann0}%
\end{equation}
which is identical in form to the Boltzmann (thermodynamic) entropy in Eq.
(\ref{Boltzmann_S}) for an isolated body in equilibrium, except that the
current formulation has been extended to an isolated body out of equilibrium;
see also Eq. (\ref{Boltzmann_S_Extended}). The only requirement is that all
microstates in $\mathbf{m}_{0}\equiv\mathbf{m}_{0}(\mathbf{Z}_{0}(t))$ are
equally probable. The statistical entropy in this case becomes a state function.

Applying the above formulation to a macrostate characterized by a given
$\mathbf{X}_{0}$ and consisting of microstates $\left\{  \overline{m}_{\alpha
}\right\}  ,$ forming the set $\overline{\mathbf{m}}_{0}\equiv\mathbf{m}%
_{0}(\mathbf{X}_{0}),$ with probabilities $\left\{  \overline{p}_{\alpha
}(t)\right\}  $, we find that
\begin{equation}
\mathcal{S}_{0}(\mathbf{X}_{0},t)\equiv-%
%TCIMACRO{\tsum \limits_{\alpha=1}^{W_{0}(\mathbf{X}_{0})}}%
%BeginExpansion
{\textstyle\sum\limits_{\alpha=1}^{W_{0}(\mathbf{X}_{0})}}
%EndExpansion
\overline{p}_{\alpha}(t)\ln\overline{p}_{\alpha}(t),%
%TCIMACRO{\tsum \limits_{\alpha=1}^{W_{0}(\mathbf{X}_{0})}}%
%BeginExpansion
{\textstyle\sum\limits_{\alpha=1}^{W_{0}(\mathbf{X}_{0})}}
%EndExpansion
\overline{p}_{\alpha}(t)\equiv1, \label{Conventional_Entropy0}%
\end{equation}
is the entropy of this macrostate, where $W_{0}(\mathbf{X}_{0})$ is the number
of distinct microstates $\overline{m}_{\alpha}$. It should be obvious that%
\[
W_{0}(\mathbf{X}_{0})\equiv%
%TCIMACRO{\tsum \nolimits_{\boldsymbol{\xi}_{0}(t)}}%
%BeginExpansion
{\textstyle\sum\nolimits_{\boldsymbol{\xi}_{0}(t)}}
%EndExpansion
W_{0}(\mathbf{Z}_{0}(t)).
\]
Again, under the equiprobable assumption
\[
\overline{p}_{\alpha}(t)\rightarrow\overline{p}_{\alpha\text{,eq}}%
=1/W_{0}(\mathbf{X}_{0}),\text{~~\ \ }\forall\overline{m}_{\alpha}\in
\Gamma_{0}(\mathbf{X}_{0}),
\]
$\Gamma_{0}(\mathbf{X}_{0})$\ denoting the space spanned by microstates
$\left\{  \overline{m}_{\alpha}\right\}  $, the above entropy takes its
maximum possible value%
\begin{equation}
\mathcal{S}_{0,\text{max}}(\mathbf{X}_{0},t)=\mathcal{S}_{0}(\mathbf{X}%
_{0})=\ln W_{0}(\mathbf{X}_{0}), \label{S_Boltzmann}%
\end{equation}
which is identical in value to the Boltzmann (thermodynamic) entropy in Eq.
(\ref{Boltzmann_S}) for an isolated body in equilibrium. The maximum value
occurs at $t=\tau_{0\text{eq}}$. It is evident that%
\begin{equation}
\mathcal{S}_{0}[\mathbf{Z}_{0}(t),t]\leq\mathcal{S}_{0}[\mathbf{Z}_{0}%
(t)]\leq\mathcal{S}_{0}(\mathbf{X}_{0}). \label{Entropy_Inequalities0}%
\end{equation}

The anticipated identification of nonequilibrium thermodynamic entropy with
$\mathcal{S}_{0}$ under some restrictions allows us to identify $\mathcal{S}%
_{0}$ as the \emph{statistical entropy} formulation of the thermodynamic
entropy. From now on, we will refer to the general entropy in Eq.
(\ref{Gibbs_Formulation}) in terms of microstate probabilities as the
\emph{time-dependent Gibbs formulation} of the entropy or simply the
\emph{Gibbs entropy}, and will not make any distinction between the
statistical and thermodynamic entropies. Accordingly, we will now use the
regular symbol $S_{0}$ for $\mathcal{S}_{0}$ throughout the work, unless
clarity is needed.

We will refer to $S_{0}(\mathbf{Z}_{0}(t))$ in terms of microstate number
$W_{0}(\mathbf{Z}_{0}(t))$ in Eq. (\ref{S_Boltzmann0})as the
\emph{time-dependent Boltzmann formulation }of the entropy or simply the
Boltzmann entropy,\cite{Lebowitz} whereas $S_{0}(\mathbf{X}_{0})$ in Eq.
(\ref{S_Boltzmann}) represents the equilibrium (Boltzmann) entropy. It is
evident that the Gibbs formulation in Eqs. (\ref{Gibbs_Formulation}) and
(\ref{Conventional_Entropy0}) supersedes the Boltzmann formulation in
Eqs.(\ref{S_Boltzmann0}) and (\ref{S_Boltzmann}), respectively, as the former
contains the latter as a special limit. However, it should be also noted that
there are competing views on which entropy is more
general.\cite{Lebowitz,Ruelle} We believe that the above derivation, being
general, makes the Gibbs formulation more fundamental. The continuity of
$S_{0}(\mathbf{Z}_{0},t)$ follows directly from the continuity of $p_{\alpha
}(t)$. The existence of the statistical entropy $S_{0}(\mathbf{Z}_{0},t)$
follows from the observation that it is bounded above by $\ln W_{0}%
(\mathbf{Z}_{0})$ and bounded below by $0$, see Eq. (\ref{S_Boltzmann0}).

It should be stressed that $\mathcal{W}$ is not the number of microstates of
the $\mathcal{N}$ replicas; the latter is given by $[W_{0}(\mathbf{Z}%
_{0})]^{\mathcal{N}}$. Thus, the entropy in Eq.
(\ref{Ensemble_entropy_Formulation}) should not be confused with the Boltzmann
entropy, which would be given by $\mathcal{N}\ln W_{0}(\mathbf{Z}_{0})$. It
should be mentioned at this point that Boltzmann uses the combinatorial
argument to obtain the entropy of a gas, see Eq. (\ref{Boltzmann_S_1}),
resulting in an expression similar to that of the Gibbs entropy in Eq.
(\ref{Gibbs_S}) except that the probabilities appearing in his formulation
represents the probability of various discrete states of a particle, and
should not be confused with the microstate probabilities used here; see Sec.
\ref{Marker_Jaynes}. The approach of Boltzmann is \emph{limited} to that of an
ideal gas only and is not general as it neglects the correlations present due
to the interactions between particles.\cite{Jaynes,Lebowitz} On the other
hand, our approach is valid for any system with any arbitrary interactions
between particles as all microstates in the collection are \emph{independent}.

\subsubsection{System in a Medium and
Quasi-independence\label{marker_Open system}}

Using the above formulation of $S_{0}(\mathbf{Z}_{0}(t),t)$, we have
determined\cite{Gujrati-II} the statistical formulation of the entropy for a
system $\Sigma$, which is a small but macroscopically large part of
$\Sigma_{0}$; see Fig. \ref{Fig.Sys}. It is assumed that the system and the
medium are \emph{quasi-independent} so that $S_{0}(t)$ can be expressed as a
sum of entropies $S(t)$ and $\widetilde{S}(t)$ of the system and the medium,
respectively:
\begin{equation}
S_{0}(t)=S(t)+\widetilde{S}(t). \label{Entropy Sum}%
\end{equation}
The two statistical entropies are given by an identical formulation
\begin{equation}
S(t)=-%
%TCIMACRO{\tsum \limits_{k}}%
%BeginExpansion
{\textstyle\sum\limits_{k}}
%EndExpansion
p_{k}(t)\ln p_{k}(t),\ \ \ \widetilde{S}(t)=-%
%TCIMACRO{\tsum \limits_{k}}%
%BeginExpansion
{\textstyle\sum\limits_{k}}
%EndExpansion
\widetilde{p}_{k}(t)\ln\widetilde{p}_{k}(t), \label{Entropies}%
\end{equation}
respectively. Here, $m_{k}$ with probability $p_{k}$ denotes a microstate of
$\Sigma$ and $\widetilde{m}_{k}$ with probability $\widetilde{p}_{k}$ that of
the medium. In the derivation,\cite{Gujrati-II} we have neither assumed the
medium nor the system to be in internal equilibrium; only quasi-independence
is assumed. The above formulation of statistical entropies will not remain
valid if the two are not quasi-independent. The same will also be true of the
thermodynamic entropies.

For the system to be in internal equilibrium, its statistical entropy $S(t)$
must be maximized under the constraints imposed by the medium. The constraints
are on the average values of the state variables:
\[
\mathbf{Z}(t)=\sum_{k}\mathbf{Z}_{k}p_{k},
\]
where $\mathbf{Z}_{k}$\ is the value of $\mathbf{Z}$ in $m_{k}$. The condition
for internal equilibrium is obtained by varying $p_{k}$ without changing the
microstates, i.e. $\mathbf{Z}_{k}$.\ Using the Lagrange multiplier technique,
it is easy to see that the condition for this in terms of the Lagrange
multipliers whose definitions are obvious is
\begin{equation}
-\ln p_{k}=\lambda_{0}+\boldsymbol{\lambda\cdot}\mathbf{Z}_{k};
\label{index_i}%
\end{equation}
the Lagrange multipliers are the same for all microstates and the scalar
product is over the elements in the set $\mathbf{Z}_{k}$. It now follows that%
\begin{equation}
S=\lambda_{0}+\boldsymbol{\lambda\cdot}\mathbf{Z}(t), \label{General Entropy}%
\end{equation}
using the same scalar product as above. It is now easy to identify the
Lagrange multipliers by observing that
\begin{equation}
dS=\boldsymbol{\lambda\cdot}d\mathbf{Z.} \label{General Entropy Differential}%
\end{equation}
Comparing this relation with the Gibbs fundamental relation for the system,
which follows from Eq. (\ref{Gibbs_Fundamental_Extended}) when applied to the
system, we find%
\[
\lambda_{p}\boldsymbol{\equiv}\left(  \partial S/\partial Z_{p}\right)  .
\]
Accepting this identification now allows us to conclude that the statistical
entropy $S(t)$ in Eq. (\ref{Entropies}) is no different than the
nonequilibrium thermodynamic entropy of the same system in internal
equilibrium but in a medium. A special case of such a system is the
(equilibrium) canonical ensemble of Gibbs. This proves (2) mentioned in Sect.
\ref{Marker_Probabilities}. \ In equilibrium, the Lagrange multipliers
associated with the internal variables vanish and Eq.
(\ref{General Entropy Differential}) reduce to%
\begin{equation}
dS=\boldsymbol{\lambda\cdot}d\mathbf{X.} \label{General Entropy Differential0}%
\end{equation}
The significance of $\lambda_{0}$\ is quite obvious. In internal equilibrium,
it is given by
\[
\lambda_{0}=S-\left(  \partial S/\partial\mathbf{Z}\right)  \boldsymbol{\cdot
}\mathbf{Z}(t).
\]
Moreover, as the nonequilibrium entropy in internal equilibrium is a state
function, it can in principle be measured or calculated by integrating Eq.
(\ref{General Entropy Differential}). Therefore, its value can be compared
with the statistical entropy. The above identification in Eq.
(\ref{General Entropy Differential}) then proves (3).

If the thermodynamic entropy is not a state function, it cannot be measured or
computed. Thus, while the statistical entropy can be computed in principle in
all cases, as shown below explicitly, there is no way to compare its value
with the thermodynamic entropy in all cases. Thus, no comment can be made
about their relationship in general. We merely conjecture that as the two
entropies are the same when the thermodynamic entropy is a state function, it
is no different from its statistical analog even when it is not a state function.

\subsection{Disjoint Sample Space (Component
Confinement)\label{Marker_Disjoint Space}}

The consideration of dynamics\ resulting in the simple connectivity of the
sample (or phase) space has played a pivotal role in developing the kinetic
theory of gases,\cite{Boltzmann0,Lebowitz} where the interest is at high
temperatures.\cite{Landau,Gujrati-Residual,Gujrati-Symmetry,Gujrati-Poincare}
As dynamics is very fast here, it is well known that the ensemble entropy
agrees with its temporal formulation. However, at low temperatures, where
dynamics becomes sluggish as in a glass,\cite{Gujrati-book,Palmer} the system
can be \emph{confined} into disjoint components.

Sample (or phase) space confinement at a phase transition such as a liquid-gas
transition is well known in equilibrium statistical
mechanics.\cite{Landau,Gujrati-Residual,Gujrati-Symmetry} It also occurs when
the system undergoes symmetry breaking such as during magnetic transitions,
crystallizations, etc. But confinement can also occur under nonequilibrium
conditions, when the observational time scale $\tau_{\text{obs}}$\ becomes
shorter than the equilibration time $\tau_{\text{eq}}$%
,\cite{Gujrati-book,Palmer} such as for glasses, whose behavior and properties
have been extensively studied.

The issue has been recently considered by us,\cite{Gujrati-Symmetry} where
only energy as an observable was considered. The discussion is easily extended
to the present case when confinement occurs for whatever reasons into one of
the thermodynamically significant number of disjoint components $\Gamma
_{0\lambda},\lambda=1,2,3\cdots,\mathcal{C}$, each component corresponding to
the same set $\mathbf{Z}_{0}$ or $\mathbf{Z}$ (we suppress the dependence for
simplicity), depending on whether the body is isolated or not. Such a
situation arises, for example, in Ising magnets at the ferromagnetic
transition., where the system is either confined to $\Gamma_{0+}$ with
positive magnetization or $\Gamma_{0-}$ with negative magnetization. Even a
weak external magnetic field $\left\vert H\right\vert \rightarrow0$, that we
can \emph{control} as an observer, will allow the system to make a choice
between the two parts of $\Gamma_{0}$. It just happens that in this case
$\mathcal{C}=2$ and is thermodynamically insignificant.

The situation with glasses or other amorphous materials is very
different.\cite{Palmer} In the first place, $\Gamma_{0}$\ is a union of
\emph{thermodynamically significant} number $\mathcal{C}\sim e^{cN}%
,c>0,$\ disjoint components. In the second place, there is no analog of a
symmetry breaking field. Therefore, there is no way to prepare a sample in a
given component $\Gamma_{0\lambda}$. Thus, the samples will be found in all
different components. Taking into consideration disjointness of the components
generalizes the number of configurations in Eq. (\ref{Combinations}) to%
\[
\mathcal{W}_{0}\mathcal{\equiv N}_{0}!/%
%TCIMACRO{\tprod \limits_{\lambda,\alpha_{\lambda}}}%
%BeginExpansion
{\textstyle\prod\limits_{\lambda,\alpha_{\lambda}}}
%EndExpansion
\mathcal{N}_{0\alpha_{\lambda}}(t)!,
\]
where $\mathcal{N}_{0\alpha_{\lambda}}$ denotes the number of sample in the
microstate $m_{\alpha_{\lambda}}$ in the $\lambda$-th component. In terms of
$p_{\alpha_{\lambda}}=\mathcal{N}_{0\alpha_{\lambda}}(t)/\mathcal{N}_{0}$,
this combination immediately leads to%
\begin{equation}
\mathcal{S}_{0}(t)\equiv\ln\mathcal{W}_{0}/\mathcal{N}_{0}=-%
%TCIMACRO{\tsum \nolimits_{\lambda}}%
%BeginExpansion
{\textstyle\sum\nolimits_{\lambda}}
%EndExpansion%
%TCIMACRO{\tsum \nolimits_{\alpha_{\lambda}}}%
%BeginExpansion
{\textstyle\sum\nolimits_{\alpha_{\lambda}}}
%EndExpansion
p_{\alpha_{\lambda}}(t)\ln p_{\alpha_{\lambda}}(t), \label{S_Component}%
\end{equation}
for the statistical entropy of the system and has already been used
earlier\cite{Gujrati-Symmetry} by us; see Sec. 4.3.3 there.\ From what has
been said above, this statistical entropy is also the thermodynamic entropy of
a nonequilibrium state under component confinement for which the entropy is a
state function of $\mathbf{Z}_{0}$. Therefore, as before, we take
$\mathcal{S}_{0}$ to be the general expression of the nonequilibrium
thermodynamic entropy and use $S_{0}$ in place of $\mathcal{S}_{0}$.

Introducing%
\[
p_{\lambda}(t)\equiv%
%TCIMACRO{\tsum \nolimits_{\alpha_{\lambda}}}%
%BeginExpansion
{\textstyle\sum\nolimits_{\alpha_{\lambda}}}
%EndExpansion
p_{\alpha_{\lambda}}(t),
\]
it is easy to see\cite{Gujrati-Symmetry} that
\[
S_{0}(t)=%
%TCIMACRO{\tsum \nolimits_{\lambda}}%
%BeginExpansion
{\textstyle\sum\nolimits_{\lambda}}
%EndExpansion
p_{\lambda}(t)S_{0\lambda}(t)+S_{\mathcal{C}}(t).
\]
Here, the entropy of the component $\Gamma_{0\lambda}$ in terms of the reduced
microstate probability $\widehat{p}_{\alpha_{\lambda}}\equiv p_{\alpha
_{\lambda}}/p_{\lambda}$ is%
\begin{equation}
S_{0\lambda}(t)=-%
%TCIMACRO{\tsum \nolimits_{\alpha_{\lambda}}}%
%BeginExpansion
{\textstyle\sum\nolimits_{\alpha_{\lambda}}}
%EndExpansion
\widehat{p}_{\alpha_{\lambda}}(t)\ln\widehat{p}_{\alpha_{\lambda}}(t)
\label{S_Single_Componet}%
\end{equation}
so that the first contribution is its average over all components. The second
term is given by%
\begin{equation}
S_{\mathcal{C}}(t)=-%
%TCIMACRO{\tsum \nolimits_{\lambda}}%
%BeginExpansion
{\textstyle\sum\nolimits_{\lambda}}
%EndExpansion
p_{\lambda}(t)\ln p_{\lambda}(t), \label{S_Residual}%
\end{equation}
and represents the component entropy. It is this entropy that is related to
the residual entropy\cite{Gujrati-Residual} in disordered systems. The same
calculation for a system in a medium will result in an identical formulation
for the entropy as in Eq. (\ref{S_Component}) except that the sum is over
components and microstates of the system. \ \ 

\section{1-dimensional ideal Gas:\ A Model Entropy Calculation}

We consider a gas of non-interacting identical structureless particles with no
spin, each of mass $m$, confined to a $1$-dimensional box of initial size
$L=L_{\text{in}}$ with impenetrable walls (infinite potential well).
Initially, the gas is in thermodynamic \emph{equilibrium} with a medium at
fixed temperature $T_{0\text{in}}$ and pressure\ $P_{0\text{in}}$. The gas is
then isolated by disconnecting it from the medium. In time, the isolated gas
expands, may be in a nonequilibrium fashion. We wish to calculate its entropy
as a function of the box size $L(t)$. \ 

As there are no interactions between the particles, the wavefunction $\Psi$
for the gas is a product of individual particle wavefunctions $\psi$. Thus, we
can focus on a single particle to study the nonequilibrium behavior of the
gas.\cite{GujTyler,GujBoyko} The simple model of a particle in a box has been
extensively studied in the literature but with a very different emphasis.
\cite{Bender,Doescher,Stutz} The particle only has non-degenerate eigenstates
whose energies are determined by $L=\alpha L_{\text{in}}$, $\alpha>1$, and \ a
quantum number $k$. We use the energy scale $\varepsilon_{1}=\pi^{2}\hbar
^{2}/2mL_{\text{in}}^{2}$ to measure the energy of the eigenstate so that%
\begin{equation}
\varepsilon_{k}(L)=k^{2}/\alpha^{2}; \label{particle-microstate energy}%
\end{equation}
the corresponding eigenfunctions are given by%
\begin{equation}
\psi_{k}(x)=\sqrt{2/L}\sin(k\pi x/L),\ \ k=1,2,3,\cdots.
\label{eigenfunctions}%
\end{equation}
The pressure generated by the eigenstate on the walls is given by
\cite{Landau-QM}
\begin{equation}
P_{k}(L)\equiv-\partial\varepsilon_{k}/\partial L=2\varepsilon_{k}(L)/L.
\label{particle-microstate presure}%
\end{equation}
In terms of the eigenstate probability $p_{k}(t)$, the average energy and
pressure are given by
\begin{subequations}
\label{particle energy pressure}%
\begin{align}
\varepsilon(t,L)  &  \equiv%
%TCIMACRO{\tsum \nolimits_{k}}%
%BeginExpansion
{\textstyle\sum\nolimits_{k}}
%EndExpansion
p_{k}(t)\varepsilon_{k}(L),\label{particle energy}\\
P(t,L)  &  \equiv%
%TCIMACRO{\tsum \nolimits_{k}}%
%BeginExpansion
{\textstyle\sum\nolimits_{k}}
%EndExpansion
p_{k}(t)P_{k}(L)=2\varepsilon(t,L)/L. \label{particle pressure0}%
\end{align}
The entropy follows from Eq. (\ref{Gibbs_Formulation}) and is given for the
single particle case by%
\end{subequations}
\[
s(t,L)\equiv-%
%TCIMACRO{\tsum \nolimits_{k}}%
%BeginExpansion
{\textstyle\sum\nolimits_{k}}
%EndExpansion
p_{k}(t)\ln p_{k}(t).
\]
The time dependence in $\varepsilon(t)$\ or $P(t)$ is due to the time
dependence in $p_{k}$ and $\varepsilon_{k}(L)$. Even for an isolated system,
for which $\varepsilon$ remains constant, $p_{k}$ cannot remain constant when
the gas is not in equilibrium if $L$ is held fixed after expansion. This
follows directly from the second law\cite{Gujrati-Symmetry} and creates a
conceptual problem because the eigenstates are mutually orthogonal and there
can be no transitions among them to allow for a change in $p_{k}$.

\subsection{Chemical Reaction Approach}

A way to change $p_{k}$ in an isolated system is to require the presence of
some stochastic interactions, whose presence allows for transitions among
eigenstates.\cite{Gujrati-Symmetry} As these transitions are happening within
the system, we can treat them as "chemical reactions" between different
eigenstates\cite{DeDonder,deGroot,Prigogine} by treating each eigenstate $k$
as a chemical species. During the transition, these species undergoes chemical
reactions to allow for the changes in their probabilities.

We follow this analogy further and extend the traditional
approach\cite{DeDonder,deGroot,Prigogine} to the present case. For the sake of
simplicity, our discussion will be limited to the ideal gas in a box; the
extension to any general system is trivial. Therefore, we will use microstates
instead of eigenstates in the following to keep the discussion general. Let
there be $N_{k}(t)$ particles in the $k$th microstate at some instant $t$ so
that
\[
N=%
%TCIMACRO{\tsum \nolimits_{k}}%
%BeginExpansion
{\textstyle\sum\nolimits_{k}}
%EndExpansion
N_{k}(t)
\]
at all times, and $p_{k}(t)=N_{k}(t)/N$. We will consider the general case
that also includes the case in which final microstates refer to a box size
$L^{\prime}$ different from its initial value $L$. Let us use $A_{k}$ to
denote the reactants (initial microstates) and $A_{k}^{\prime}$ to denote the
products (final microstates). For the sake of simplicity of argument, we will
assume that transitions between microstates is described by a single chemical
reaction, which is expressed in stoichiometry form as
\begin{equation}%
%TCIMACRO{\tsum \nolimits_{k}}%
%BeginExpansion
{\textstyle\sum\nolimits_{k}}
%EndExpansion
a_{k}A_{k}\longrightarrow%
%TCIMACRO{\tsum \nolimits_{k}}%
%BeginExpansion
{\textstyle\sum\nolimits_{k}}
%EndExpansion
a_{k}^{\prime}A_{k}^{\prime}. \label{General Reaction}%
\end{equation}
Let $N_{k}$ and $N_{k}^{\prime}$ denote the population of $A_{k}$ and
$A_{k}^{\prime}$, respectively, so that $N=%
%TCIMACRO{\tsum \nolimits_{k}}%
%BeginExpansion
{\textstyle\sum\nolimits_{k}}
%EndExpansion
N_{k}=%
%TCIMACRO{\tsum \nolimits_{k}}%
%BeginExpansion
{\textstyle\sum\nolimits_{k}}
%EndExpansion
N_{k}^{\prime}$. Accordingly, $p_{k}(t)=N_{k}(t)/N$ for the reactant and
$p_{k}(t+dt)=N_{k}^{\prime}(t)/N$ for the product. The single reaction is
described by a single extent of reaction $\xi$ and we have
\[
d\xi(t)\equiv-dN_{k}(t)/a_{k}(t)\equiv dN_{k^{\prime}}^{\prime}%
(t)/a_{k^{\prime}}^{\prime}(t)\text{ \ \ for all }k,k^{\prime}.
\]
It is easy to see that the coefficients satisfy an important relation%
\[%
%TCIMACRO{\tsum \nolimits_{k}}%
%BeginExpansion
{\textstyle\sum\nolimits_{k}}
%EndExpansion
a_{k}(t)=%
%TCIMACRO{\tsum \nolimits_{k}}%
%BeginExpansion
{\textstyle\sum\nolimits_{k}}
%EndExpansion
a_{k}^{\prime}(t),
\]
which reflects the fact that the change $\left\vert dN\right\vert $ in the
reactant microstates is the same as in the product microstates. The
\emph{affinity} in terms of the chemical potentials is given by%
\[
A(t)=%
%TCIMACRO{\tsum }%
%BeginExpansion
{\textstyle\sum}
%EndExpansion
a_{k}(t)\mu_{A_{k}}(t)-%
%TCIMACRO{\tsum }%
%BeginExpansion
{\textstyle\sum}
%EndExpansion
a_{k}^{\prime}(t)\mu_{A_{k}^{\prime}}(t),
\]
and will vanish only in "equilibrium," i.e. only when $p_{k}$' s attain their
equilibrium values. Otherwise, $A(t)$ will remain non-zero. It acts as the
thermodynamic force in driving the chemical
reaction.\cite{DeDonder,deGroot,Prigogine} But we must wait long enough for
the reaction to come to completion, which happens when $A(t)$ and $d\xi/dt$
both vanish. The extent of reaction $\xi$ is an example of an internal
variable. For the ideal gas under consideration, there does not seem to be any
other internal variable as particles have no internal structures. In the
following, we will assume only one internal variable $\xi(t)$.%

%TCIMACRO{\FRAME{ftbpFU}{3.4714in}{2.6662in}{0pt}{\Qcb{The calculated
%equilibrium (continuous) and nonequilibrium (broken) entropies per particle
%for an ideal gas in a box as a function of the expansion box length $L$. The
%nonequilibrium state is the result of a sudden expansion from the initial
%state corresponding to $L=1$ and $T_{0}=4.$ The energy of the gas remains
%constant in the sudden expansion. As expected, the nonequilibrium entropy lies
%below the equilibrium entropy. In time, the former will increase to the latter
%as the gas equilibrates. }}{\Qlb{Fig.Entropy}}{entropy_comparison.eps}%
%width 3.4714in;  height 2.6662in;  depth 0pt;  original-width 6.2604in;
%original-height 4.7971in;  cropleft "0";  croptop "1";  cropright "1";
%cropbottom "0";  filename 'Entropy_Comparison.EPS';file-properties "XNPEU";}%
%}}%
%BeginExpansion
\begin{figure}
[ptb]
\begin{center}
\includegraphics[
height=2.6662in,
width=3.4714in
]%
{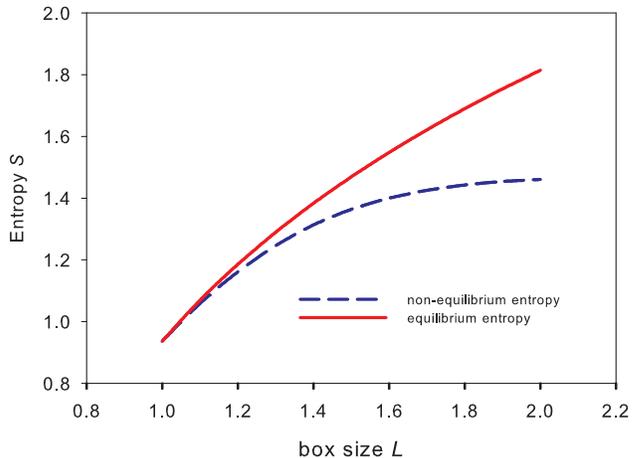}%
\caption{The calculated equilibrium (continuous) and nonequilibrium (broken)
entropies per particle for an ideal gas in a box as a function of the
expansion box length $L$. The nonequilibrium state is the result of a sudden
expansion from the initial state corresponding to $L=1$ and $T_{0}=4.$ The
energy of the gas remains constant in the sudden expansion. As expected, the
nonequilibrium entropy lies below the equilibrium entropy. In time, the former
will increase to the latter as the gas equilibrates. }%
\label{Fig.Entropy}%
\end{center}
\end{figure}
%EndExpansion

\subsection{Free (Sudden) Expansion of the Box}

The box expands as a function of time, which need not be quasi-static
(extremely slow) so there is no reason to assume that the gas remains in
equilibrium after expansion. The entropy of the gas per particle can be
obtained by calculating $s(t,L)=-%
%TCIMACRO{\tsum \nolimits_{k}}%
%BeginExpansion
{\textstyle\sum\nolimits_{k}}
%EndExpansion
p_{k}(t)\ln p_{k}(t)$ for the particle under consideration. Henceforth, we
will call $s(t)$ the entropy of the particle, which shares the property that
the irreversible entropy change $d_{\text{i}}s(t)$ will never be negative. All
the above discussion about the chemical reaction is easily translated to the
study of a particle in box without any change. The change $d_{\text{i}}%
p_{n}(t)$ is caused by the transitions between different eigenstates.

We consider the gas in \emph{equilibrium} at some initial temperature
$T_{0\text{in}}$ in a box of length $L_{\text{in}}$, which we take to be
$L_{\text{in}}=1$. This is obtained by keeping the box in a medium of
temperature $T_{0\text{in}}$. The corresponding microstate probabilities
follow the Boltzmann law ($\beta_{\text{in}}\equiv1/T_{0\text{in}}$):%
\[
p_{k\text{,eq}}(\beta_{\text{in}},L_{\text{in}})=\exp(-\beta_{\text{in}%
}\varepsilon_{k}(L_{\text{in}}))/Z_{0}(\beta_{\text{in}},L_{\text{in}}),
\]
where $Z_{0}(\beta_{\text{in}},L_{\text{in}})\equiv%
%TCIMACRO{\tsum \nolimits_{k}}%
%BeginExpansion
{\textstyle\sum\nolimits_{k}}
%EndExpansion
\exp(-\beta_{\text{in}}\varepsilon_{k}(L_{\text{in}}))$ denotes the
equilibrium partition function. The energy per particle in this gas is denoted
by $\varepsilon_{\text{in}}$ obtained by replacing $p_{k}(\beta,L)$ by
$p_{k\text{,eq}}(\beta_{0},L_{\text{in}})$ in Eq.
(\ref{particle energy pressure}); the corresponding pressure is $P_{0\text{in}%
}=2\varepsilon_{\text{in}}/L_{\text{in}}$. The equilibrium entropy can be
obtained by using $p_{k\text{,eq}}(\beta_{0},L_{\text{in}})$ in $S$ given in
Eq. (\ref{Entropies}).\ The initial temperature $T_{0\text{in}}$ for
$L_{\text{in}}=1$ is taken to be $T_{0\text{in}}=4$ so that the initial energy
$\varepsilon_{\text{in}}\approx2.786$.

We now consider equilibrium states having the same initial thermodynamic
energy $\varepsilon_{\text{in}}$ for different values of $L$, even though the
eigenstate energies and their Boltzmann probabilities vary with $L$. The
corresponding equilibrium entropy $s_{\text{eq}}(t)$\ as a function of $L$ is
shown in Fig. \ref{Fig.Entropy} by the continuous curve. Since the energy is
constant, the product $P_{0\text{eq}}(L)L=2\varepsilon_{\text{in}}$ is also
constant; see Eq. (\ref{particle energy pressure}). Thus, $P_{0\text{eq}}(L)$
is a decreasing function of $L$. From the slope of the upper curve in Fig.
\ref{Fig.Entropy} which decreases with $L$, we also conclude that
$P_{0\text{eq}}(L)/T_{0\text{eq}}(L)$ is a decreasing function of $L$. Thus,
$T_{0\text{eq}}(L)L$ is an increasing function of $L$. However, our
calculation to be presented elsewhere \cite{GujTyler} shows that
$T_{0\text{eq}}(L)$ is also an increasing function of $L$. The eigenstates for
a box of size $L$ are given in Eq. (\ref{eigenfunctions}).

We now consider nonequilibrium states. For this, we isolate the box from its
medium and consider its free expansion as it expands suddenly from
$L_{\text{in}}$ to a new size $\ L>L_{\text{in}}$. Because of its isolation,
its energy remains $\varepsilon_{\text{in}}$ during this expansion. As the
expansion is sudden, the initial eigenfunctions $\psi_{l\text{in}}(x)$ for
$L_{\text{in}}$ have no time to change, but are no longer the eigenfunctions
of the new size $L$; the latter are given by $\psi_{k}(x)$ in Eq.
(\ref{eigenfunctions}) for $L$. However, $\psi_{l\text{in}}(x)$ can be
expanded in terms of $\psi_{k}(x)$ as a sum over $k$. We call this the quantum
superposition principle. The corresponding expansion coefficients $b_{kl}$ are
easily seen to be\cite{Bender}%
\[
b_{kl}(L,L_{\text{in}})=\frac{2l\alpha^{3/2}(-1)^{l}}{\pi(k^{2}-\alpha
^{2}l^{2})}\sin(\frac{k\pi}{\alpha}).
\]
Using $b_{kl}$\ and $p_{l}(\beta_{\text{in}},L,L_{\text{in}})$, we can
determine the probability $p_{k}(\beta_{\text{in}},L,L_{\text{in}})$ for the
$k$th microstate in the new box, which allows us to determine all
thermodynamic averages for the new box. We have checked that the new
probabilities add to $1$ and that the (average) energy after the free
expansion is equal to $\varepsilon_{\text{in}}$ to within our computational
accuracy. Thus, $\Delta\varepsilon=0$ in the sudden expansion. This is
consistent with the fact that the gas does no external work and that no
external heat is exchanged.

Despite this, the free expansion is spontaneous once the confining walls have
moved. Therefore, the (thermodynamic) entropy of the gas must increase in this
process in accordance with the second law. We use $p_{k}(\beta_{\text{in}%
},L,L_{\text{in}})$\ to evaluate the nonequilibrium statistical entropy, which
is shown by the dashed curve in Fig. \ref{Fig.Entropy}. The significance of
this curve is as follows: Choose a particular value $L$ in this graph. Then,
the nonequilibrium entropy for this $L$ is given by numerically evaluating the
sum
\[
s(\beta_{\text{in}},L,L_{\text{in}})=-%
%TCIMACRO{\tsum \nolimits_{k}}%
%BeginExpansion
{\textstyle\sum\nolimits_{k}}
%EndExpansion
\ p_{k}(\beta_{\text{in}},L,L_{\text{in}})\ln p_{k}(\beta_{\text{in}%
},L,L_{\text{in}}).\
\]
This is the entropy after the sudden expansion from the initial state at
$L_{\text{in}}=1$ and follows from the quantum superposition principle.
Evidently, this entropy is higher that the initial equilibrium entropy
$s_{\text{eq}}(\beta_{\text{in}},L_{\text{in}})$. It is also obvious that this
entropy has a memory of the initial state at $L_{\text{in}}=1$ and
$T_{0\text{in}}=4$. Therefore, it does not represent the equilibrium entropy.
If we now wait at the new value of $L$, the isolated gas in the new box will
relax to approach its equilibrium state in which its nonequilibrium entropy
will gradually increase until it becomes equal to its value on the upper curve.

\section{1-d Tonks Gas:\ A simple Continuum Model}

A careful reader would have realized by this time that the proposed entropy
form in Eq. (\ref{Gibbs_Formulation}) is not at all the same as the standard
classical formulation of entropy, such as for the ideal gas, which can be
negative at low temperatures or at high pressures. The issue has been
discussed elsewhere\cite{Guj-Fedor} but with a very different perspective.
Here, we visit the same issue from a very different perspective that allows us
to investigate if and how the entropy in continuum models is related to the
proposed entropy in this work. For this, we turn to a very simple continuum
model in classical statistical mechanics: the Tonks gas,\cite{Tonks,Thompson}
which is an athermal model and contains the ideal gas as a limiting case when
the rod length $l$ vanishes. We will simplify the discussion by considering
the Tonks gas in one dimension. The gas consists of $r$\ impenetrable rods,
each of length $l$ lying along a line of length $L$. We will assume $r$ to be
fixed, but allow $l$ and $L$ to change with the state of the system, such as
its pressure.. The configurational entropy per rod determined by the
configurational partition function is found to be\cite{Thompson}%
\begin{equation}
s_{\text{c}}=\ln[e(v-l)], \label{Tonks_S}%
\end{equation}
where $v$ is the "volume" available per rod $L/r$. Even though the above
result is derived for an equilibrium Tonks gas, it is easy to see that the
same result also applies for the gas in internal equilibrium. The only
difference is that the parameters in the model are also functions of internal
variables now.

The entropy vanishes when $v=l+1/e$ and becomes negative for all $v<l$.
Indeed, it diverges to $-\infty$ in the incompressible limit $v=l$. This is
contrary to the Boltzmann approach in which the entropy is determined by the
number of microstates (cf. Eq. (\ref{Boltzmann_S})) or the Gibbs approach (cf.
Eq. (\ref{Gibbs_Formulation})) and can never be negative. Can we reconcile the
contradiction between the continuum entropy and the current statistical formulation?

We now demonstrate that the above entropy for the Tonks gas is derivable from
the current statistical approach under some approximation, to be noted below,
by first considering a lattice model for the Tonks gas and then taking its
continuum limit. It is in the lattice model can we determine the number of
microstates. In a continuum, this number is always \emph{unbounded }(see below
also). For this we consider a $1$-d lattice $\Lambda_{\text{f}}$ with
$N_{\text{f}}$ sites; the lattice spacing, the distance between two
consecutive sites, is given by $\delta$. We take $N_{\text{f}}>>1$ so that
$L_{\text{f}}=(N_{\text{f}}-1)\delta\approx N_{\text{f}}\delta$ is the length
of the the lattice $\Lambda_{\text{f}}$. We randomly select $r$ sites out of
$N_{\text{f}}$. The number of ways, which then represents the number of
configurational microstates, is given by%
\begin{equation}
W_{\text{c}}=N_{\text{f}}!/r!(N_{\text{f}}-r)!. \label{Tonks_Microstates}%
\end{equation}
After the choice is made, we replace each selected site by $\lambda+1$
consecutive sites, each site representing an atoms in a rod, to give rise to a
rod of length $l\equiv\lambda\delta$. It is clear that $\delta$\ also changes
with the state of the system. The number of sites in the resulting lattice
$\Lambda$ is
\[
N=N_{\text{f}}+r\lambda
\]
so that the length of $\Delta$ is given by $L=(N-1)\delta\approx N\delta$
since $N>>1$. We introduce the number densities $\varphi_{\text{f}%
}=r/N_{\text{f}}$, $\rho_{\text{f}}=r/N_{\text{f}}\delta\approx r/L_{\text{f}%
}$ and $\rho=r/N\delta\approx r/L$. A simple calculation shows that $S=\ln
W_{\text{c}}$ is given by%
\[
S=-N_{\text{f}}[\rho_{\text{f}}\delta\ln\rho_{\text{f}}\delta+\ln
(1-\rho_{\text{f}}\delta)-\rho_{\text{f}}\delta\ln(1-\rho_{\text{f}}\delta)].
\]
This result can also be obtained by taking the athermal entropy for a
polydisperse polymer solution a Bethe lattice\cite{Gujrati-Monodisperse} by
setting the coordination number $q$ to be $q=2$. We now take the continuum
limit $\delta\rightarrow0$ \ for \emph{fixed} $\rho_{\text{f}}$ and $\rho$,
that is \emph{fixed} $L_{\text{f}}$ and $\rho L$, respectively. In this limit,
$\ln(1-\rho_{\text{f}}\delta)\approx-\rho_{\text{f}}\delta$, and
$\rho_{\text{f}}\delta\ln(1-\rho_{\text{f}}\delta)\approx-(\rho_{\text{f}%
}\delta)^{2}$. Use of these limits in $S$ yields%
\begin{equation}
S=-r\ln(e/\rho_{\text{f}}\delta)\rightarrow\infty. \label{Cont_S}%
\end{equation}
The continuum limit of the entropy from the Boltzmann approach has resulted in
a diverging entropy regardless of the value of $\rho_{\text{f}}$%
,\cite{Guj-Fedor} a well known result. By introducing an arbitrary
\emph{constant} $a$ with the dimension of length, we can rewrite $S$ as%
\begin{equation}
S/r=-\ln(e/\rho_{\text{f}}a)+\ln(\delta/a), \label{Cont_Manipulation}%
\end{equation}
in which the first term remains finite in the continuum limit, and the second
term contains the divergence. The diverging part, although explicitly
independent of $\rho_{\text{f}}$, still depends on the state of the gas
through $\delta$, and cannot be treated as a constant unless we assume
$\delta$\ to be independent of the state of the gas. It is a common practice
to approximate the lattice spacing $\delta$ as a constant. In that case, the
diverging term represents a constant that can be subtracted from $S/r$.
Recognizing that $1/\rho_{\text{f}}=v-l$, we see that the first term in Eq.
(\ref{Cont_Manipulation}) is nothing but the entropy of \ the Tonks gas in Eq.
(\ref{Tonks_S}) for the arbitrary constant $a=1$. However, this equivalence
only occurs in the state independent constant-$\delta$ approximation.

As the second term above has been discarded, the continuum entropy
$s_{\text{c}}$ also has \emph{no} simple relationship with the number ($\geq
1$) of microstates in the continuum limit, which means that the continuum
entropy cannot be identified as the Boltzmann entropy in Eq.
(\ref{S_Boltzmann}). To see this more clearly, let us focus on the centers of
mass of each rod, which represent one of the $r$ sites that were selected in
$\Lambda_{\text{f}}$. Each of the $k$ sites $x_{k}$, $k=1,2,\cdots,r$, is free
to move over $L_{\text{f}}$. The \emph{adimensional} volume $\left\vert
\Gamma_{\text{f}}\right\vert $, also called the probability and denoted by $Z$
by Boltzmann,\cite{Note-Boltzmann,Lebowitz} of the corresponding phase space
$\Gamma_{\text{f}}$ is $L_{\text{f}}^{r}/a^{r}$. However, contrary to the
conventional wisdom,\cite{Lebowitz} $\ln\left\vert \Gamma_{\text{f}%
}\right\vert $ does not yield $s_{\text{c}}$. The correct expression is given
by the Gibbs-modified adimensional volume $\left\vert \Gamma_{\text{f}%
}\right\vert /r!$, i.e.
\[
\frac{1}{r!a^{r}}L_{\text{f}}^{r}.
\]
The presence of $r!$ is required to restrict the volume due to
indistinguishability of the rods \`{a} la Gibbs. For large $r$, this quantity
correctly gives the entropy $s_{\text{c}}$. However, this quantity is not only
not an integer, it also cannot be \emph{always} larger than or equal to unity,
as noted above.

\section{Semi-Classical Approximation in Phase Space for $S$%
\label{Marker_SemiClassical}}

The analog of a quantum microstate in classical statistical mechanics is
normally obtained by recognizing that in the \emph{adiabatic approximation},
each small phase space cell of volume element $dx_{\text{c}}(=dpdq$ in terms
of generalized coordinates $q$ and momenta $p)$ of size $h^{s}$ corresponds to
a microstate,\cite{Landau,Rice} where $s$ is the degrees of freedom of the
system. The latter follows from the Bohr-Sommerfeld quantization rule for a
periodic motion. The adiabatic approximation requires the parameters
characterizing the system to vary extremely slowly. We will assume one such
parameter $\lambda$ (such as the volume $V$) so that \cite{Landau-Mech}
$\tau\overset{.}{\lambda}<<\lambda,$ where $\tau$ is the period of oscillation
for constant $\lambda$; the system would be isolated in the latter case. In
the above approximation, the energy of the system will vary very slowly and
can be taken to be constant over a period of oscillation. The action taken
over the closed path for the constant value of $\lambda$ and the energy is
quantized \cite{Landau-QM}:%
\[
I=%
%TCIMACRO{\toint }%
%BeginExpansion
{\textstyle\oint}
%EndExpansion
pdq=(n+1/2)h.
\]
This observation is the justification of the above cell size of a classical
microstate. Thus, the number of "classical" microstates is given by%
\[
W=\Delta\Gamma/h^{s},
\]
where $\Delta\Gamma$ is the phase space volume corresponding to the system.
This allows us to divide the phase space into $W$ cells, index by $k$, of
volume $dx_{k}=dx_{\text{c}}$ and "centered" at $x_{k}$ at some time which we
call the initial time $t=0$. We will denote the evolution of the cell at time
$t$ by the location of its "center" $x_{k}(t)$ and its volume element
$dx_{k}(t)$. In terms of the distribution function $f(x,t)$ in the phase
space, the $k$th microstate probability therefore is given by%
\begin{equation}
p_{k}(t)\equiv f(x_{k},t)dx_{k}(t). \label{f-probability}%
\end{equation}
Evidently,
\begin{equation}%
%TCIMACRO{\tsum \nolimits_{k}}%
%BeginExpansion
{\textstyle\sum\nolimits_{k}}
%EndExpansion
f(x_{k},t)dx_{k}(t)=1 \label{f-Normalization}%
\end{equation}
at all times. The entropy and the average of any observable $O(x,t)$ of the
system are given by%
\begin{subequations}
\begin{align}
S(t)  &  \equiv-%
%TCIMACRO{\tsum \nolimits_{k}}%
%BeginExpansion
{\textstyle\sum\nolimits_{k}}
%EndExpansion
f(x_{k},t)dx_{k}(t)\ln[f(x_{k},t)dx_{k}(t)],\label{Discrete_S}\\
\overline{O}(t)  &  \equiv%
%TCIMACRO{\tsum \nolimits_{k}}%
%BeginExpansion
{\textstyle\sum\nolimits_{k}}
%EndExpansion
O(x_{k},t)f(x_{k},t)dx_{k}(t), \label{Discrete_O}%
\end{align}
the sum being over all\ $W$ microstates. While it is easy to see that
continuum analogs for Eqs. (\ref{f-Normalization}) and (\ref{Discrete_O}) are
easily identified, this is not so for the entropy in Eq.\ (\ref{Discrete_S}%
).\cite{Jaynes-prob} However, it should be obvious that $S_{f}$ in Eq.
(\ref{Sf_S}) cannot be a candidate for the statistical entropy $S(t).$

It is well known \cite{Landau-QM} that the system in the adiabatic limit
remains in the same quantum state. For Hamiltonian dynamics, the conservation
of the phase space cell volume under evolution ensures $dx_{k}=dx_{\text{c}}$
for each cell so that $\overset{.}{x}=0$. This results in the
incompressibility of the phase space. In equilibrium, $f(x_{k},t)$ is not only
uniform but also constant, and we conclude from Eq. (\ref{f-Normalization})
that this value is $f=1/\Delta\Gamma$. Accordingly, $p_{k}=1/W$ in equilibrium
as expected and we obtain the equilibrium entropy $S_{\text{eq}}=\ln W$.

The situation is far from clear for nonequilibrium states. As the example of
expansion of the box shows, the system is no longer restricted to be in the
same microstate, which means that the microstate energy is no longer a
constant and the phase space trajectory is no longer closed. Thus, the
suitability of the Bohr-Sommerfeld quantization is questionable, and care must
be exercised to identify the microstates. We will adopt the following
prescription. We consider some equilibrium state (uniform $f(x)$) of the
isolated system to identify the cell volume $dx_{\text{c}}=$ $h^{s}$. Once the
identification has been done, we will no longer worry about its relationship
with $h^{s}$, and only deal with the cells. We then follow the evolution of
each cell in time in a nonequilibrium process during which $\overset{.}{x}=0$
may not hold. Thus the volume of each cell may no longer be constant. The
process may also result in changes in $f(x_{k},t)$.\cite{Holian} Indeed, it is
quite possible that $f$ diverges at the same time that $dx_{\text{c}}$
vanishes.\cite{Hoover} However, their product, which determines the microstate
probability, must remain strictly bounded and $\leq1$. In particular, as the
cell volume shrinks to zero, $f$ must diverge to keep the product bounded.
Thus, the divergence\cite{Ramshaw,Hoover0} of $f$ alone does not imply that
the entropy diverges to negative infinity.

This is easily seen by the following 1-d damped oscillator, the standard
prototype of a dissipative system:\cite{Landau-Mech} $\overset{..}{x}%
+2\kappa\overset{.}{x}+\omega_{0}^{2}x=0$ with a (positive) damping
coefficient, which is chosen such that $\kappa>\omega_{0}$ just for
simplicity. We have the case of aperiodic damping. We will only consider the
long time behavior. It is easy to see that in this limit ($\kappa^{\prime
}=\kappa+\sqrt{\kappa^{2}-\omega_{0}^{2}}$)%
\end{subequations}
\[
dx_{\text{c}}\sim\exp(-2\kappa^{\prime}t),\ f\sim\exp(+2\kappa^{\prime}t),
\]
and their product remains bounded, as expected.

\section{Jaynes Revisited\label{Marker_Jaynes}}

Boltzmann\cite{Boltzmann} provides the following alternative expression of the
entropy\cite{Cohen,Boltzmann} in terms of a single particle probability
$p_{i}^{(1)}$ for the particle to be in the $i$th state:%
\begin{equation}
S_{\text{B}}^{(1)}=-N%
%TCIMACRO{\tsum \nolimits_{i}}%
%BeginExpansion
{\textstyle\sum\nolimits_{i}}
%EndExpansion
p_{i}^{(1)}\ln p_{i}^{(1)}, \label{Boltzmann_S_1}%
\end{equation}
not to be confused with that in Eq. (\ref{Boltzmann_S}). Boltzmann is only
interested in the maximum entropy, which occurs when all states are equally
probable. In this case,
\[
S_{\text{B, max}}^{(1)}=N\ln w
\]
where $w$ is the number of possible states of a single particle in the gas. In
general, particles are not independent due to interactions and number of
possible states $W<w^{N}$. Accordingly, maximum Gibbs entropy $S_{\text{ max}%
}$ per particle is \emph{less} than the corresponding equiprobable Boltzmann
entropy $S_{\text{B, max}}^{(1)}$. However, Jaynes\cite{Jaynes} gives a much
stronger results, see his Eq. (5):%
\[
S<S_{\text{B}}^{(1)}.
\]
The equality occurs only if there are no interactions between the particles,
as we have asserted above.\ \ \ 

\section{Summary and Discussion\label{marker_Summary}}

Recognizing that there does not exists a first principles statistical
formulation of nonequilibrium thermodynamic entropy for an isolated system in
terms of microstate probabilities, we have attempted to fill in the gap. We
use a formal approach (frequentist interpretation of probability) by extending
the equilibrium ensemble of Gibbs to a nonequilibrium ensemble, which is
nothing but a large number $\mathcal{N}_{0}$ of samples of the thermodynamic
system under consideration. Accordingly, we refer to the ensemble as a sample
space. The formal approach enables us to evaluate the combinatorics for a
given set of microstate probabilities. The resulting statistical entropy is
independent of the number of samples and depends only on the probabilities as
is seen from Eqs. (\ref{Gibbs_Formulation}) and (\ref{S_Component}). Thus, the
use of a large number of samples is merely a formality and is not required in
practice. We have shown that in equilibrium, the statistical entropy is the
same as the equilibrium thermodynamic entropy: $\mathcal{S}_{0}(\mathbf{X}%
_{0})=S_{0}(\mathbf{X}_{0})$. But we have also shown that the statistical
entropy is equal to the nonequilibrium thermodynamic entropy, provided the
latter is a state function of the nonequilibrium state variables
$\mathbf{Z}_{0}(t)$: $\mathcal{S}_{0}(\mathbf{Z}_{0}(t))=S_{0}(\mathbf{Z}%
_{0}(t))$. We cannot make any comment about the relationship between
$\mathcal{S}_{0}(\mathbf{Z}_{0}(t),t)$ and $S_{0}(\mathbf{Z}_{0}(t),t)$ for
the simple reason that there is no way to measure or calculate a non-state
function $S_{0}(\mathbf{Z}_{0}(t),t)$. We should remark here that the standard
approach to calculate nonequilibrium entropy is to use the classical
nonequilibrium thermodynamics\cite{deGroot} or its variant, which treats the
entropy at the local level as a state function.

Some readers may think that our statistical formulation is no different than
that used in the information theory. We disagree. For one, there is no concept
of internal variables $\boldsymbol{\xi}_{0}$ in the latter theory. Because of
this, our approach allows us to consider three levels of description so that
we can consider three different entropies $S(\mathbf{Z}_{0}(t),t),S_{0}%
(\mathbf{Z}_{0}(t))$ and $S_{0}(\mathbf{X}_{0})$ satisfying the inequalities
in Eq. (\ref{Entropy_Inequalities0}). The information theory can only deal
with two levels of entropies. There is also no possibility of a residual
entropy in the latter.

For an isolated system in internal equilibrium ($p_{\alpha}=p$ for $\forall
m_{\alpha}$), just a single sample will suffice to determine the entropy as
samples are \emph{unbiased}. The entropy in this case is no different than the
"entropy" $-\ln p$ of a single sample:\cite{Gujrati-Symmetry,Lebowitz}%
\[
S(t)=(-p\ln p)W_{0}=-\ln p=\ln W_{0},
\]
where $W_{0}$ represents $W_{0}(\mathbf{Z}_{0}(t))$ or $W_{0}(\mathbf{X}_{0}%
)$. However, this simplicity is lost as soon as the system is not in internal
equilibrium. Here, one must consider averaging over all microstates.\ 

Changes in microstate probabilities result in changes in the entropy. There
are two ways probabilities can change within an isolated system, both of them
being irreversible in nature. One cause of changes is due to the quantum
nature as seen in the sudden expansion of the box. Here, the parameter
$\lambda$ ($=L$) changes non-adiabatically and creates irreversibility. The
resulting irreversible change in the entropy for the $1$-$d$ gas has been
calculated and shown by the lower curve in Fig. \ref{Fig.Entropy}. The other
cause of probability changes is due to the "chemical reaction" going on among
the microstates that brings about equilibration in the system. The
corresponding irreversible rise in the entropy for the gas is shown by the
difference between the two curves in Fig. \ref{Fig.Entropy}. The interaction
of a body with its medium can also result in the changes in microstate
probabilities, and has been considered elsewhere.\cite{Gujrati-Heat-Work}

We consider the continuum analog of the statistical formulation of entropy and
show that the standard formulation, $S_{f}$ in Eq. (\ref{Sf_S}), is not a good
candidate of the nonequilibrium entropy. It is then argued that the divergence
of $f$ in some cases, see the discussion above, makes $S_{f}$ diverge to
$-\infty$, even though the statistical entropy remains finite and positive.
Thus, $S_{f}$ cannot be equated with our statistical formulation, a
generalization of the Gibbs formulation. We suggest that our statistical Gibbs
formulation can be applied to any nonequilibrium state.


\begin{thebibliography}{99}
\bibitem {Note-Boltzmann}The number of combinations in Eq. (35) on p. 56 in
Boltzmann\cite{Boltzmann0} is denoted by $Z$, but it is not the number of
microstates. The two become the same only when $Z$ is maximized as discussed
on p. 58.
\end{thebibliography}
\end{document}